\documentclass[aps,prd,twocolumn,showpacs,groupedaddress,floatfix]{revtex4}

\usepackage{graphicx}
\usepackage{longtable}
\DeclareGraphicsExtensions{.eps, .eps, .jpg, .png}

\begin{document}

\preprint{hep-th/0502035\\ January 2005}
\date{\today}

\title{Gravitational Shock Waves and Their Scattering in Brane-Induced Gravity}

\author{Nemanja Kaloper} 
\affiliation{Department of Physics, University of California,
Davis, CA 95616 }

\begin{abstract}
In this paper, following hep-th/0501028, we present a detailed
derivation and discussion of  the exact gravitational field
solutions for a relativistic particle localized on a tensional
brane in brane-induced gravity. Our derivation yields the metrics
for both the normal branch and the self-inflating branch Dvali,
Gabadadze and Porrati (DGP) braneworlds. They generalize the 4D
gravitational shock waves in de Sitter space, and so we compare
them to the corresponding 4D General Relativity solution and to
the case when gravity resides only in the 5D bulk, and there are
no brane-localized graviton kinetic terms. We write down the
solutions in terms of two-variable hypergeometric functions and
find that at short distances the shock wave profiles look exactly
the same as in 4D Minkowski space, thus recovering the limit one
expects if gravity is to be mediated by a metastable, but
long-lived, bulk resonance. The corrections far from the source
differ from the long distance corrections in 4D de Sitter space,
coming in with odd powers of the distance. We discuss in detail
the limiting case on the self-inflating branch when gravity is
modified exactly at de Sitter radius, and energy can be lost into
the bulk by resonance-like processes. Finally, we consider
Planckian scattering on the brane, and find that for a
sufficiently small impact parameter it is approximated very
closely by the usual 4D description.
\end{abstract}
\pacs{11.25.-w, 11.25.Mj, 98.80.Cq, 98.80.Qc \hfill
hep-th/0502035}

\maketitle


\section{Introduction}

Among many recent braneworld proposals, a particularly interesting
idea called brane-induced gravity was put forth by Dvali,
Gabadadze and Porrati (DGP) \cite{dgp}. In this approach the
gravitational force between masses on the brane should arise from
the exchange of an unstable bulk graviton resonance, which falls
apart at very large scales, instead of a stable zero-mode graviton
governed by the usual four-dimensional (4D) General Relativity
(GR). Far from sources the law of gravity is modified, and changes
from the 4D to a fully higher-dimensional one. This is similar to
the approach of \cite{grs}, which is known to suffer from problems
with ghosts \cite{grsghosts} (see however \cite{ghosts} for a
possible cure of the ghost-induced maladies). The brane-induced
gravity setup of DGP may be free of ghosts in perturbation theory
\cite{dgp}, and so it is an interesting arena to explore
modifications of gravity at large distances. In na\"ive
perturbation theory, the modifications appear already at the level
of linearized theory, where one finds that the tensor structure of
the graviton-matter coupling changes from its form in 4D GR
\cite{dgp}. Indeed, around flat space, a small perturbation
$h_{\mu\nu}$ is sourced by $T_{\mu\nu} - \alpha \eta_{\mu\nu}
T^\lambda{}_\lambda$, where $\alpha = 1/3$ instead of $1/2$,
because of the extra scalar polarization of the massive graviton.
Since this coupling does not depend on the graviton mass, the
factor $1/3$ would appear to persist even as $m_g \rightarrow 0$.
This is an example of van Dam-Veltman-Zakharov (vDVZ)
discontinuity \cite{vvdz}, which could modify the predictions for
the Solar system tests of general relativity. However, Vainshtein
has argued \cite{nonflat} that the na\"ive perturbation theory for
massive gravity breaks down in the limit $m_g \rightarrow 0$
because it is a double expansion in both $G_N$ and $1/m_g$.
Instead, the argument goes, one needs to expand around the curved
background created by the source mass, where the curvature
generated by the source ``screens" the scalar graviton, so that
the modified perturbation theory smoothly reduces to conventional
GR as $m_g \rightarrow 0$. Basically, it is necessary to take into
account the non-linear nature of gravity to properly restore gauge
invariance which is required to ensure the decoupling of the
massive modes as $m_g \rightarrow 0$. A similar behavior has also
been noticed in non-abelian gauge theories \cite{nonflat}.

The issue of strong coupling in gravity with IR modifications, and
in particular its implications for DGP models have been studied in
\cite{strongcouplings,nimags,lpr,rubakov,giga,nira}. The works
\cite{lpr,rubakov,giga,nira} specifically address the longitudinal
graviton and its matter coupling, and suggest that brane extrinsic
curvature may indeed play the role of a coupling controller. When
a source mass is placed on the brane, it bends the brane. The
brane extrinsic curvature which locally responds to the source may
in turn help to tame the perturbation theory \cite{nira}. This
interplay between the source-induced background curvature and the
scalar graviton coupling may thus be a key ingredient of the
``gravity filter" of \cite{dgp}. It is therefore essential to
explore in detail this interplay of the curvature and the coupling
strength beyond perturbation theory in order to clarify the status
of the effective 4D theory. However, the DGP equations are very
difficult to solve exactly for compact sources localized on the
brane \cite{schwarz,tanaka}.

In the previous paper \cite{kallet} we have presented the first
explicit example of an exact solution for a localized particle in
DGP models. We have obtained the gravitational field for a
relativistic particle on a tensional brane in 5D, which
generalizes 4D gravitational shock waves
\cite{pirani,aichsexl,thooft,gabriele,devega,barrabes} in de
Sitter space \cite{hota,pogri,kostas}. In this paper we follow up
with a detailed derivation of the shock wave metrics for both the
normal branch and the self-inflating branch variants of DGP
braneworlds \cite{dsbranes}. Focusing on the extreme relativistic
limit, we find that the brane-localized terms are indeed a very
efficient ``gravity filter". At short distances, the wave profile
behaves exactly like in 4D GR \cite{aichsexl,thooft}. The
deviations appear only far from the source, where they come as odd
powers of the distance from the source in the units of the
cosmological horizon length, as opposed to the 4D GR corrections
in de Sitter space, which only come as even powers. We also look
at gravitational scattering of relativistic particles in DGP
braneworlds, and find that to the leading order it behaves the
same as in the 4D case.

We note that our solutions do not provide a direct answer to the
conundrum of strong couplings of the scalar graviton. However, we
do find that the scalar graviton field is not turned on in these
metrics. This would not come as a surprise in a weakly-coupled
perturbation theory, where the scalar graviton will not be sourced
by ultra-relativistic particles since to the leading order their
source vanishes, $T^{\mu}{}_\mu = 0$. In other words, in the
relativistic limit, one finds that the conformal symmetry is
restored, and the solution is form-invariant under boosts. Hence
the gravitational shock wave of 4D GR is also a solution in
Brans-Dicke theory. In a more relevant case of very fast particles
with non-zero rest mass, the corrections from the mass to the
relativistic wave profile should be suppressed by the source
mass-to-momentum ratio $m/p \ll 1$. Our results show that
decoupling of the scalar graviton persists in the relativistic
limit in DGP, and so the dangerous strongly coupled mode
identified in perturbation theory does not destroy the solutions
at large momenta after all. This suggests that even for the
particles with nonzero rest mass the gravitational field may be
under control. Indeed, since local physics in DGP obeys the usual
4D diffeomorphism invariance, we can always pick a very
fast-moving observer to explore the field of a massive source, and
transform all of the relevant physics to her rest frame. In her
frame, thanks to Relativity Principle, the source mass would
appear to move with a very high speed ${\tt v}$, and therefore its
gravitational field should be well approximated by our shock wave
solutions. For a sufficiently fast observer, the corrections due
to the mass should be suppressed by the powers of $m/p =
\sqrt{1/{\tt v}^2-1}$. This suggests that if organized as an
expansion in powers of $m/p$, perturbation theory may be under
control (however, by itself this may not help with any ghost
infestations on the self-inflating branch \cite{lpr}).

The paper is organized as follows. In the next section we
construct the background solutions for a tensional brane without
particle excitations on it. They are de Sitter vacua of 5D DGP. In
section III we present a detailed derivation of the gravitational
shock waves, and show that they can be written in terms of
two-variable hypergeometric functions. We outline a spectacular
new non-perturbative channel for production of bulk gravitons on
the self-inflating branch, when gravity is modified at exactly the
brane de Sitter radius, in section IV, and consider the limiting
form of the solutions at short distances, as well as several
special cases in section V. We then turn to the problem of
gravitational scattering of relativistic particles in section VI,
and summarize in section VII.

\section{de Sitter vacua}

Brane-induced gravity models \cite{dgp} are given by a bulk action
with metric kinetic terms in both the bulk and on the brane. In
the case of one extra dimension,
\begin{eqnarray}
S &=& \int d^5x \sqrt{g_5} \, \frac{M^3_5}{2} \, R_5 - \int d^4x
\sqrt{g_4} \, M^3_5 K \nonumber \\
&+&  \int d^4x \sqrt{g_4} \Bigl(\frac{M^2_4}{2} \, R_4 - \lambda -
{\cal L}_{\rm matter} \Bigr) \, . \label{action}
\end{eqnarray}
Here $g_{AB}$ and $g_{\mu\nu} = \partial_\mu X^A
\partial_\nu X^B g_{AB}$ are the 5D metric and the induced 4D metric on
the brane, respectively, $R_5$ and $R_4$ are their Ricci tensors,
$K = g^{AB} K_{AB}$ is the usual Gibbons-Hawking term necessary in
space-times with boundaries (defined as the trace of the extrinsic
curvature of the brane, for variational principles see e.g.
\cite{chambre,shtanov}), and $\lambda$ and ${\cal L}_{\rm matter}$
the brane tension and matter Lagrangian (which we separate from
each other explicitly). The indices $\{A,B\}$ and $\{\mu,\nu\}$
are bulk and 3-brane world-volume indices, and the gravitational
couplings are controlled by the bulk and brane Planck scales,
$M_4$ and $M_5$, respectively.

Varying this action away from the brane one finds the
gravitational field equations in the bulk, which in the case of
empty bulk admit locally 5D Minkowski solution. The brane dynamics
can then be incorporated via the modified Israel junction
condition \cite{giga,dsbranes}, which in the case of brane-induced
gravity in addition to the brane extrinsic curvature also receives
corrections from the intrinsic curvature as well, arising from the
variation of $R_4$. This junction condition can be viewed as the
equation of motion of the brane, defining the world-volume the
brane sweeps in the locally Minkowski bulk. In the case of empty
bulk, one can readily transform to the rest frame of the brane by
going to the Gaussian normal coordinates for the 5D metric,
$ds_5^2 = dw^2 + g_{\mu\nu}(w,x) dx^\mu dx^\nu$, and incorporate
the brane equations of motion as $\delta$-function terms in the
bulk field equations, taking a very compact form. In 5D, those
field equations are \cite{dgp}
\begin{equation}
M^3_5 G_5^A{}_B  + M^2_4 G_4^\mu{}_\nu \delta^A_\mu \delta^\nu_B
\delta(w) = - T^\mu{}_\nu  \delta^A_\mu \delta^\nu_B \delta(w)\, ,
\label{fieldeqs}
\end{equation}
where $G_5^A{}_B$, $G_4^\mu{}_\nu$ are bulk and 3-brane-localized
Einstein tensors, and $T^{\mu}{}_\nu$ is the brane stress-energy,
respectively. In this coordinate system one can easily orbifold
the bulk, by imposing a $Z_2$ symmetry $ w \rightarrow - w$ around
the brane situated at $w=0$.

For a brane with tension $\lambda \ne 0$ and without any localized
or distributed matter sources, $T^{\mu}{}_{\nu} = - \lambda
\delta^\mu{}_\nu$. The corresponding solutions describe de Sitter
vacua on the brane. The symmetries of the brane-bulk system then
dictate the form of the metric, which must be given by the warped
4D de Sitter space in a locally flat bulk \cite{dsbranes}
\begin{equation}
ds^2_5 = (1- \epsilon H|w|)^2 ds^2_{4dS} + dw^2 \, .
\label{background}
\end{equation}
The replacement $w \rightarrow |w|$ implements the orbifolding
procedure. Aside from the parameter $\epsilon$, which can take
values $\pm 1$ in (\ref{background}), this metric is identical to
the 5D version \cite{kalinde} of the inflating
Vilenkin-Ipser-Sikivie domain wall \cite{vis}. In what follows we
shall use the static patch de Sitter metric,
\begin{equation}
ds^2_{4dS} = -(1-H^2r^2) dt^2 + \frac{dr^2}{(1 - H^2 r^2)} + r^2
d\Omega_2 \, , \label{des} \end{equation}
instead of the spatially flat one utilized in \cite{kalinde}. The
space-time of (\ref{background}) is a 4D de Sitter hyperboloid in
a 5D Minkowski bulk. Here $\epsilon$ arises because we can retain
either the interior $\epsilon = + 1$ or the exterior $\epsilon =
-1$ of the hyperboloid after orbifolding, thanks to the brane
intrinsic curvature $R_4$. The junction conditions relate the 4D
curvature and the brane tension as \cite{dsbranes}
\begin{equation}
H^2 + \epsilon \frac{2 M^3_5}{M_4^2} H = \frac{\lambda}{3 M^2_4}
\, . \label{hubble}
\end{equation}
The simplest way to derive this equation is to substitute the
metric (\ref{background}) into the field equations
(\ref{fieldeqs}), recall that away from the brane
(\ref{background}) is just a different parameterization of the
locally flat Minkowski metric, and compute the terms proportional
to $\delta(w)$, which in this case arise from the double
$w$-derivatives in the bulk terms on the LHS in (\ref{fieldeqs}),
and from the brane curvature terms. Then matching them with the
tension source yields (\ref{hubble}). We will revisit this
derivation in more detail in section III.

\begin{figure}[thb]
\vskip.2cm
\centerline{\includegraphics[width=0.25\hsize,width=0.25\vsize,angle=0]{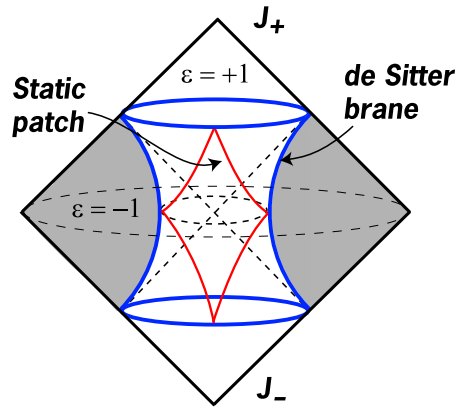}}
\caption{{\small Penrose diagram of a de Sitter brane in a flat 5D
bulk. We suppress one angular coordinate on $S^2$ on the brane and
depict another, so that the diagram looks like two cones glued
along their base, with the hyperboloid of the brane world volume
inscribed inside the cones. In the case of the normal branch, in
the bulk one retains the interior of the brane world-volume,
denoted with $\epsilon = +1$. For the self-inflating branch, one
retains the exterior, marked $\epsilon = -1$. The static patch on
the brane is the interior of the diamond-shaped region on the
brane world-volume.} \label{fig:three}}
\end{figure}

Focusing on the solutions with $H >0$ (because $H<0$ cases are
their $PT$ transforms), on the normal branch defined by the choice
$\epsilon = +1$, we see that the equation (\ref{hubble}) reduces
to the 4D Friedman equation in the limit $M_5 \ll M_4$, $3H^2
\simeq \lambda/M^2_{4}$. In this case in the bulk we keep the
interior of the hyperboloid. It has finite volume and so the
perturbative 4D graviton exists. The brane curvature terms $\sim
M^2_{4} \int d^4x \sqrt{g_4} R_4/2$ suppress the couplings of the
$m_g > 0$ graviton KK modes, producing the 4D effective theory
when $M_5 \ll M_4$. In Fig. \ref{fig:three}. this case corresponds
to keeping the unshaded region of the bulk, i.e. the interior of
the de Sitter hyperboloid.

On the other hand, on the self-inflating branch, defined by
$\epsilon = -1$, at low tensions $\lambda \ll  12 M^6_5/M^2_4$,
from the eq. (\ref{hubble}) one finds $H \sim 2 M^3_5/M_4^2$. In
this case in the bulk one keeps the exterior of the hyperboloid
that opens up to infinity, corresponding to the shaded region in
Fig. \ref{fig:three}. This bulk volume is infinite, and so there
is no perturbative 4D graviton. The effective theory arises only
from the exchange of the bulk resonance.

We note that this solution resembles Witten's ``bubble of nothing"
\cite{witten} (see also \cite{gregory}, see Fig.
\ref{fig:three}.), which first collapses from infinite size to a
minimum, and bounces back to infinity, gobbling up the bulk along
the way. However now the bubble wall carries finite energy
density. This may help protect the bulk from being ``eaten" by
``nothing" too quickly. Namely, the probability of nucleating a
``bubble of nothing" is given by the exponential of the negative
action of its Euclidean extension, ${\cal P} \simeq \exp(-S_E)$.
If we take the self-inflating branch solutions (\ref{background})
and glue them to a Euclidean hemisphere at the minimum radius to
model bubble nucleation, we find that the corresponding action is
\begin{equation}
S_E = \frac{V_4}2 (\lambda + 16 M^3_5H)  \, , \label{euclact}
\end{equation}
where $V_4$ is the volume of the sphere of radius $H^{-1}$ (for a
simple calculation of the action with Gibbons-Hawking terms see
e.g. \cite{chambre,bent}). In the limit $\lambda \ll 12
M^6_5/M^2_4$, $M_5 \ll M_4$, only big bubbles with a large
Euclidean action, $S_E \sim (M_4/M_5)^6 \gg 1$, are nucleated.
Therefore, if the tension and the bulk Planck scale are somehow
fixed to be much smaller than the 4D Planck scale $M_4$ (see e.g.
\cite{numbers} for the discussion of scales in brane-induced
gravity) the nucleation rate of ``bubbles of nothing" may be
sufficiently slow to render the self-inflating branch long-lived.

\section{Derivation of Shock Waves}

We now turn to the construction of the gravitational shock wave
fields generated by relativistic particles moving on the brane in
DGP \cite{kallet}. Imagine that a photon is moving on a tensional
brane, with a momentum $p = 2\pi \nu$. Because it is a
relativistic particle trapped on the brane, it moves along a null
geodesic of the brane-induced metric. By the Principle of
Equivalence, its momentum generates gravitational field, since it
contributes to the stress-energy tensor. There are two approaches
for generating the gravitational field solutions. One could start
with a linearized solution for a massive particle at rest, and
boost it to relativistic speed, simultaneously taking the limit $m
\rightarrow 0$ such that $m \cosh \gamma = p$ remains constant.
This is how the solutions in 4D were originally found
\cite{pirani,aichsexl}. The procedure is completely analogous to
the method of generating the electro-magnetic potential of a
charged massless particle (for a nice review, see
\cite{barrabes}). The same procedure has also been applied to the
massless particles moving in maximally symmetric spaces
\cite{hota,pogri} and Randall-Sundrum braneworld models
\cite{aref,roberto}. The reason why this procedure works as a
method of generating {\it exact} solutions of gravitational field
equations is that if we expand the exact metric for a massive
source in powers of the mass $m$, after a boost to relativistic
speeds the quadratic and higher powers of $m$ vanish because there
is only one factor of the boost parameter $\cosh \gamma$ in the
metric, which is compensated by one power of $m$ when $\cosh
\gamma \rightarrow \infty$, $m \rightarrow 0$ and $m \cosh \gamma
= p = {\rm const}$. Hence the linearized solution becomes exact.
Essentially, in the extreme relativistic limit the solution
becomes scale invariant, and the non-linear effects which are
controlled by the fixed ratio $m/M_4$ drop out.

The previous discussion on boosting linearized solutions
describing point masses to generate exact solutions of
gravitational field equations illuminates the nature of the
solution as a planar shock wave. Because of the infinite boosting
the field lines of the gravitational field become confined to a
plane perpendicular to the direction of motion. In terms of the
null coordinate $u$ which parameterizes the world-line of the
particle $v$, the field experiences a jump at $u=0$ encoding
causal propagation of the source. When the particle flies by an
observer, its gravitational field perturbs the observer at
precisely that instant \cite{thooft}. This property allowed Dray
and 't Hooft to develop a very elegant cut-and-paste technique
\cite{thooft} to find solutions, which they applied to
asymptotically flat 4D backgrounds. Higher-dimensional examples
were considered in \cite{gabriele,devega}. This technique was
later applied to general 4D GR backgrounds by Sfetsos
\cite{kostas}. To circumvent the need for a detailed form of the
linearized solution for a localized mass in DGP, we will apply a
variant of this cut-and-paste technique to DGP. We start by
rewriting (\ref{background}) in suitable null coordinates,
defining $u = \frac{1}{H} \sqrt{\frac{1-Hr}{1+Hr}} \exp(Ht)$ and
$v = \frac{1}{H} \sqrt{\frac{1-Hr}{1+Hr}} \exp(-Ht)$. We also
introduce a new bulk coordinate $|z| = -\frac{1}{\epsilon H} \ln(1
- \epsilon H |w|)$. It simplifies evaluating the curvature tensors
with a conformal map as in \cite{conformal}. Note that $\delta(w)
= \delta(z)$ because the warp factor and its derivative are equal
to unity on the brane. In terms of these coordinates, we can
rewrite the metric (\ref{background}) in the following form:
\begin{equation}
ds_5^2 = e^{-2 \epsilon H |z|} \Bigl\{ \frac{4 dudv}{(1+H^2 uv)^2}
+ (\frac{1-H^2 uv}{1+H^2 uv})^2 \frac{d\Omega_2}{H^2} + dz^2
\Bigr\} \, . \label{nullback}
\end{equation}

Suppose now that a photon moves along a null geodesic on the
brane, let's say the $v$-axis of the coordinates utilized in
(\ref{nullback}). The equation of this null geodesic is $u=0$. If
we choose the origin of this photon at infinite past as the North
Pole of the static patch, an observer located there sees a photon
streaming away along the past horizon. Following Dray and 't
Hooft, we introduce the shock wave on top of the background
(\ref{nullback}) by including a jump in the $v$ coordinate at
$u=0$ \cite{thooft}. We substitute
\begin{eqnarray}
&& v \rightarrow v + \Theta(u) f \, , \nonumber \\
&& dv \rightarrow dv + \Theta(u) df \, , \label{jump}
\end{eqnarray}
where $f$ is the shock wave profile, whose precise form is to be
determined by requiring that combining (\ref{nullback}) and
(\ref{jump}) still remains a solution of the field equations. The
wave profile $f$ depends only on the spatial coordinates
transverse to the wave, which in our case are the angles on the
2-sphere and the bulk coordinate $z$. Here $\Theta(u)$ is the
Heaviside step function. Note that in flat 4D background
(\ref{jump}) is a trivial diffeomorphism if $f = {\rm const}.$
\cite{thooft}. In that surroundings, if $df \ne 0$, the
replacement rule for $dv$ in (\ref{jump}) ensures this is not just
a diffeomorphism. Changing the coordinates to $\hat v = v +
\Theta(u) f$ we find $dv \rightarrow d \hat v - \delta(u) f du$.
Substituting $v, dv \rightarrow \hat v, d \hat v - \delta(u) f du$
in (\ref{nullback}) and dropping the carets, we obtain the
ans\"atz for the metric with the shock wave included:
\begin{eqnarray}
ds_5^2 &=& e^{-2 \epsilon H |z|} \Bigl\{ \frac{4 dudv}{(1+H^2
uv)^2} - \frac{4 \delta(u) f du^2}{(1+H^2 uv)^2} + \nonumber \\
&& ~~~~~~~~~~~~~~ + (\frac{1-H^2 uv}{1+H^2 uv})^2 \,
\frac{d\Omega_2}{H^2} + dz^2 \Bigr\} \, . \label{wave}
\end{eqnarray}
Note that with this, the ans\"atz for the induced metric on the
brane is
\begin{eqnarray}
ds_4^2 &=& \frac{4 dudv}{(1+H^2
uv)^2} - \frac{4 \delta(u) f du^2}{(1+H^2 uv)^2} + \nonumber \\
&& ~~~~~~~~~~~~~~ + (\frac{1-H^2 uv}{1+H^2 uv})^2 \,
\frac{d\Omega_2}{H^2}  \, . \label{brane}
\end{eqnarray}
Now we take the ans\"atz (\ref{wave}), (\ref{brane}) and
substitute it into the field equations (\ref{fieldeqs}), where we
also include the photon stress-energy contribution in the source
on the RHS. Thus the total stress-energy tensor with a photon
moving along $u=0$ is
\begin{equation} T^{\mu}{}_{\nu} = - \lambda \delta^\mu{}_\nu + 2
\frac{p}{\sqrt{g_5}} g_{4\,uv} \delta(\theta) \delta(\phi)
\delta(u) \delta^\mu_v \delta^u_\nu \, . \label{stress}
\end{equation}
We have chosen the coordinates on the 2-sphere such that the
photon trajectory is at $\theta=0$.

To evaluate the curvature we utilize the conformal transformation
trick as in \cite{conformal}, splitting the metric as $ds_5^2 =
\Omega^2 d\bar s_5^2$, where $\Omega = \exp(-\epsilon H |z|)$, and
\begin{eqnarray}
d\bar s_5^2 &=& \frac{4 dudv}{(1+H^2 uv)^2} - \frac{4 \delta(u) f
du^2}{(1+H^2 uv)^2} + \nonumber \\
&& ~~~~~~~~ + \, (\frac{1-H^2 uv}{1+H^2 uv})^2 \,
\frac{d\Omega_2}{H^2} + dz^2 \, . \label{confwave}
\end{eqnarray}
A straightforward albeit tedious computation, where we treat
$\delta(u)$ and its derivatives as distributions, and use the
distributional identities $u \delta(u) = 0$, $u^2 \delta^2(u) = 0$
and $f(u) \delta'(u) = - f'(u) \delta(u)$ \cite{thooft,kostas},
yields the Ricci tensor for the conformal metric (\ref{confwave}).
The stress-energy (\ref{stress}) is automatically conserved,
$\nabla_\mu T^\mu{}_\nu \propto u \delta(u) = 0$, so the matter
sector field equations are satisfied. We note that this
computation also yields automatically the components of the Ricci
tensor of the induced metric on the brane, which appear as the
subset of the curvature tensor for the conformal metric
(\ref{confwave}), because $\Omega(0) = 1$ and the coordinates in
(\ref{wave}) are Gaussian, as is obvious from comparing
(\ref{brane}) and (\ref{confwave}). The resulting non-vanishing
Ricci tensor components are
\begin{eqnarray}
\bar R_{5\,uu} &=& 2 \delta(u) \Bigl(\partial_z^2 f + H^2 (
\Delta_2 f - 4f) \Bigr) \, , \nonumber \\
R_{4\,uu} &=& 2 H^2 \delta(u) (
\Delta_2 f - 4f) \, , \nonumber \\
\bar R_{5\,uv} &=& R_{4\,uv} = \frac{6H^2}{(1+H^2uv)^2} \, , \nonumber \\
\bar R_{5\,ab} &=& R_{4\,ab} = 3 \Bigl(\frac{1-H^2uv}{1+H^2uv}
\Bigr)^2 g_{ab} \, , \label{ricciconf}
\end{eqnarray}
where $\bar R_{5\,AB}$ and $R_{4\,\mu\nu}$ are the Ricci tensors
of the metrics (\ref{confwave}) and (\ref{brane}), and $a,b$ and
$g_{ab}$ are the indices and the metric on the unit 2-sphere,
respectively. The operator $\Delta_2$ is the Laplacian on the unit
2-sphere.

Using the conformal transformation $ds_5^2 = \Omega^2 d\bar s_5^2$
which gives the relationship
\begin{eqnarray}
R_{5\,AB} &=& \bar R_{5\,AB} - 3 \bar \nabla_A \bar \nabla_B \ln
\Omega + 3 \bar \nabla_A \ln \Omega \bar \nabla_B \Omega \nonumber
\\
&& ~~~~~~~~  - \bar g_{AB} \Bigl(\bar \nabla^2 \ln \Omega + 3
(\bar \nabla \ln \Omega)^2 \Bigr) \, , \label{conftrans}
\end{eqnarray}
we finally find
\begin{eqnarray}
R_{5\,zz} &=& 8\epsilon H \delta(z) \, ,  \nonumber \\
R_{5\,uu} &=& 2 \delta(u) \Bigl(\partial_z^2 f - 3\epsilon H
\partial_{|z|}
f + H^2 (\Delta_2 f + 2f) \Bigr) \nonumber \\
&& - 8 \epsilon H f \delta(u) \delta(z) \, , \nonumber \\
R_{5\,uv} &=& \frac{4\epsilon H}{(1+H^2uv)^2} \delta(z) \, , \nonumber \\
R_{5\,ab} &=& \frac{2\epsilon }{H} \Bigl(\frac{1-H^2uv}{1+H^2uv}
\Bigr)^2 \delta(z) g_{ab} \, . \label{riccicomp}
\end{eqnarray}
Although Ricci tensor vanishes away from the brane at $z=0$ even
when the shock wave is present ($R_{5\,uu}$ in (\ref{riccicomp})
vanishes at $z \ne 0$ by virtue of the field equations, see
below), the bulk is not Minkowski any more, since the waves extend
off the brane and deform the bulk. One would find their
non-vanishing contributions away from the brane in the Riemann
tensor. Now from (\ref{ricciconf}) and (\ref{riccicomp}) we can
obtain the explicit form of the field equations (\ref{fieldeqs}).
First, it is easy to check that we can rewrite (\ref{fieldeqs}) as
\begin{eqnarray}
&& M^3_5 R_{5\,zz} + \frac{M^2_4}3 R_4 \delta(z) = \frac43 \lambda
\delta(z) \, , \label{riccieqs} \\
&& M^3_5 R_{5\,\mu\nu} + M^2_4 (R_{4\,\mu\nu} - \frac16
g_{4\,\mu\nu} R_4) \delta(z) = \nonumber \\
&& ~~~~~ = \frac{\lambda}{3} g_{4\,\mu\nu} \delta(z) + H^2 p
(g_{4\,uv})^2 \delta(\Omega) \delta(u) \delta(z) \delta^u_\mu
\delta^u_\nu \, . \nonumber
\end{eqnarray}
Here $\delta(\Omega)$ is the covariant $\delta$-function on the
unit 2-sphere, $\delta(\Omega) = \delta(\theta)
\delta(\phi)/\sqrt{g_2} = \delta(\cos \theta - 1) \delta(\phi)$,
which peaks at $\theta=0$. Substituting (\ref{ricciconf}) and
(\ref{riccicomp}) in the first of (\ref{riccieqs}) we recover just
the background equation (\ref{hubble}), linking the Hubble
parameter $H$ and the brane tension $\lambda$. The new equation
for the wave profile $f$ comes from the ${uu}$ component of
(\ref{riccieqs}). Indeed, direct substitution of (\ref{ricciconf})
and (\ref{riccicomp}) yields a {\it linear} field equation for the
wave profile:
\begin{eqnarray}
&& \frac{M^3_5}{M_4^2 H^2} \Bigl(\partial_z^2 f - 3 \epsilon H
\partial_{|z|} f + H^2( \Delta_2 f + 2f) \Bigr) + \nonumber \\
&& ~~~~~~~~~~~~~ + (\Delta_2 f + 2f) \delta(z) = \frac{2p}{M^2_4}
\delta(\Omega) \delta(z) \, . \label{feqn}
\end{eqnarray}
The remaining components of the second of (\ref{riccieqs}) again
reduce to the background equation (\ref{hubble}). Thus
(\ref{feqn}) fully determines the shock wave profile $f$. Note
that in the limit $M_5 \rightarrow 0$ the bulk derivatives
disappear, and we can factor out $\delta(z)$ to recover the
correct 4D de Sitter equation of \cite{hota,kostas}.

\begin{figure}[thb]
\vskip.2cm
\centerline{\includegraphics[width=0.25\hsize,width=0.25\vsize,angle=0]{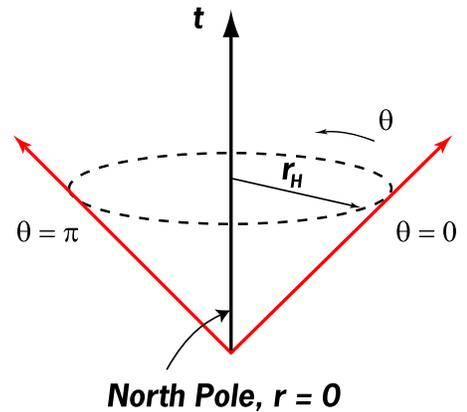}}
\caption{{\small Trajectories of two photons in the reference
frame of an observer at rest on the North Pole, at $r=0$ in the
static patch. The photons move along opposite directions on the
$S^2$, at coordinate distance $r_H = 1/H$ from the pole.  We are
explicitly depicting the polar angle $\theta$ and suppressing the
azimuthal angle $\phi$ on $S^2$.} \label{slika}}
\end{figure}

Now we can turn to solving (\ref{feqn}). It turns out that one has
to carefully implement the boundary conditions for the single
particle problem (\ref{feqn}) in order to avoid additional
singularities that would arise because the 2-sphere on which the
particle propagates is compact \cite{kostas}. These boundary
conditions are handled in the simplest way by adding an extra
particle with the same momentum $p$, running in the opposite
direction in the static patch (\ref{nullback}) \cite{hota,pogri}.
In fact, this two-source solution correctly represents the limit
of infinite boost of the Schwarzschild-de Sitter geometry
\cite{hota,pogri}. So following \cite{hota,pogri,kostas} we add an
extra term, $\frac{2p}{M^2_4} \delta(\Omega') \delta(z)$, on the
RHS of (\ref{feqn}). Here $\delta(\Omega')$ is peaked on the
opposite pole on the 2-sphere, at $\theta = \pi$ (see Fig.
\ref{slika}). Then we can return to a single source by taking this
solution and multiplying it by $\Theta(\pi/2 - \theta)$ as in
\cite{kostas}.

So let us consider eq. (\ref{feqn}) with two photons, when one of
them moves along $\theta = 0$ and the other along $\theta = \pi$:
\begin{eqnarray}
\frac{M^3_5}{M_4^2 H^2} \Bigl(\partial_z^2 f - 3 \epsilon H
\partial_{|z|} f + H^2( \Delta_2 f + 2f) \Bigr) + ~~~~~ && \nonumber \label{feqn2}\\
+ (\Delta_2 f + 2f) \delta(z) = \frac{2p}{M^2_4} \Big(
\delta(\Omega) + \delta(\Omega') \Bigr) \delta(z) \, . &&
\end{eqnarray}
Since the sources on the RHS are $\delta(\Omega) = \delta(\cos
\theta - 1) \delta(\phi)$ and $\delta(\Omega') = \delta(\cos
\theta + 1) \delta(\phi)$, they can be decomposed in terms of
spherical harmonics as
\begin{eqnarray}
\delta(\Omega) &=& \sum_{l=0}^\infty \sum_{m=-l}^l Y^*_{lm}(0,0)
\, Y_{lm}(\theta,\phi)
\, , \nonumber \\
\delta(\Omega') &=& \sum_{l=0}^\infty \sum_{m=-l}^l Y^*_{lm}(0,0)
\, Y_{lm}(\pi - \theta, \phi) \, . \label{spherharm}
\end{eqnarray}
Since we chose the photon trajectories to be along $\theta = 0$
and $\theta = \pi$, we can use the addition theorem for spherical
harmonics to replace them with Legendre polynomials $P_l(\cos
\theta)$ in the expansion for the wave profile. Indeed, from
$\sum_{m=-l}^l Y^*_{lm}(0,0) \, Y_{lm}(\theta,\phi) =
\frac{2l+1}{4\pi} P_l(\cos \theta)$ and the similar equation for
the $\theta \rightarrow \pi - \theta$ terms, and linearity of
(\ref{feqn2}), we deduce that the solutions are of the form
\begin{equation}
f = \sum_{l=0}^\infty \Bigl( f_l^{(+)}(z) P_l(\cos \theta) +
f_l^{(-)}(z) P_l(-\cos \theta) \Bigr) \, . \label{solans}
\end{equation}
Here $f_l^{(\pm)}(z)$ are the bulk wave functions; $f_l^{(+)}$ is
sourced by the photon at $\theta=0$ and $f_l^{(-)}$ by the photon
at $\theta = \pi$. By orthogonality and completeness of Legendre
polynomials, using the expansions for the $\delta$-functions from
above, the field equation (\ref{feqn2}) yields the same
differential equation for the modes $f_l^{(\pm)}(z)$:
\begin{eqnarray}
 \partial_z^2 f_l - 3 \epsilon H
\partial_{|z|} f_l + H^2(2 - l(l+1) ) f_l = ~~~~~~~~~ && \nonumber \\
 = \frac{M_4^2 H^2}{M^3_5} \Bigl( \frac{(2l+1)p}{2 \pi M^2_4}
- (2 - l(l+1))f_l \Bigr) \delta(z) \, . && \label{modesn}
\end{eqnarray}
We should interpret $\delta$-function on the RHS by pillbox
integration as a matching condition for the first derivatives of
$f_l^{(\pm)}$ on the brane, recalling also the orbifold symmetry
which enforces $f_l(-z) = f_l(z)$ .
This yields the boundary value problem for $f_l$ on the brane:
\begin{eqnarray}
&& \partial_z^2 f_l - 3 \epsilon H
\partial_{z} f_l + H^2(2 - l(l+1)) f_l = 0 \, , ~~~~~ \nonumber  \\
&& f_l(-z) = f_l(z) \, , ~~~~~ \label{bvprob} \\
&& f'_l(0) + \frac{2-l(l+1)}{\tt g} H  f_l(0) = \frac{H}{\tt g}
\frac{2l+1}{2\pi M^2_4} \, p  \, . ~~~~~ \nonumber
\end{eqnarray}
where ${\tt g} = 2 M^3_5/(M^2_4 H) = 1/(H r_c)$ (see \cite{dgp}).
Since both $f_l^{(\pm)}$ solve the same boundary value problem,
$f_l^{(+)} = f_l^{(-)} = f_l$. Because $P_l(-x) = (-1)^l P_l(x)$,
the solution will be an expansion only in even-indexed polynomials
$P_{2l}(\cos \theta)$, circumventing the unphysical 4D
singularities in $l=1$ terms like in \cite{hota,kostas}:
\begin{equation}
f = 2 \sum_{l=0}^\infty f_{2l}(z) P_{2l}(\cos \theta) \, .
\label{solseven}
\end{equation}

We still need to specify the boundary conditions for $f$ in the
limit $z\rightarrow \infty$. We do so by requiring square
integrability in the bulk, so that the solutions are localized on
the brane. To see how to implement square integrability on the
modes $f_l$, it is sufficient to consider the simpler example of a
free massless bulk scalar field, with action
\begin{equation}
S = - \int d^5 x \sqrt{g_5} (\nabla \chi)^2 \, . \label{scalact}
\end{equation}
In terms of the conformal geometry (\ref{confwave}), the action is
\begin{equation}
S = - \int dz d^4 x \sqrt{g_4} \, \Omega^3 \, (\bar \nabla \chi)^2
\, , \label{confscalact}
\end{equation}
where $\Omega = \exp(- \epsilon H |z|)$. Therefore, the norm for
the expansion of $\chi$ in terms of the 4D modes satisfies
\begin{equation}
|| \chi_\alpha ||^2 \propto \int dz \, e^{- 3\epsilon H |z|} \,
\chi_\alpha^2 \, \label{norm}
\end{equation}
We are interested in the modes of $\chi$ which are simple
exponentials of $|z|$, because from (\ref{bvprob}) it follows that
$f_l$'s are such simple exponentials. Hence let $ \chi_\alpha(x)
\propto e^{-\alpha |z|}$; then from (\ref{norm}) we conclude that
a mode will be normalizable only if $\alpha \ge (1-3\epsilon
H)/2$. Solving the differential equation in (\ref{bvprob}) we see
that the modes are of the form $\sim e^{[\pm 2l - (\mp 1 -
3\epsilon)/2] \, H |z|}$. Hence we should retain only the
sub-leading bulk mode which is square-integrable, or equivalently,
localized on the brane:
\begin{equation}
f_l = A_l e^{- [2 l +(1-3\epsilon)/2] \, H |z|} \, .
\label{asymptotic}
\end{equation}

We can now write down explicitly the series solution of
(\ref{bvprob}) which satisfies these boundary conditions.
Substituting (\ref{asymptotic}) in the last of (\ref{bvprob}) and
solving the ensuing algebraic equation for $A_l$, we find
\begin{eqnarray}
f(\Omega,z) &=& - \frac{p}{2\pi M^2_4} \sum_{l=0}^\infty
\frac{4l+1}{(2l - 1 + \frac{(1-\epsilon){\tt g}}{2})
(l+1 + \frac{(1+\epsilon){\tt g}}{4})} \nonumber \\
&& ~~~~~~~~~~~ \times e^{- [2 l +(1-3\epsilon)/2] \, H |z|}
P_{2l}(\cos \theta) \, , \label{sols}
\end{eqnarray}
As a check, in the limit ${\tt g}=0$, at $z=0$ this reduces to the
4D series solution of \cite{hota} for both $\epsilon = \pm 1$. We
can write the series (\ref{sols}) in the integral form. Defining
$\tau = e^{- H |z|}$ and $x = \cos \theta$ and factorizing the
coefficients of the expansion, we first find
\begin{eqnarray}
f(\Omega,z) = - \frac{[3 - (1-\epsilon) {\tt g}] \, p }{(3 +
\epsilon {\tt g})\, \pi M^2_4} \sum_{l=0}^\infty \frac{\tau^{2
l+(1-3\epsilon)/2}}{2l - 1 +
\frac{(1-\epsilon){\tt g}}{2}} P_{2l}(x) && \nonumber \\
- \frac{[3 + (1+\epsilon){\tt g}] \, p}{2 (3 + \epsilon {\tt g})
\pi M^2_4}  \sum_{l=0}^\infty \frac{\tau^{2 l
+(1-3\epsilon)/2}}{l+1 + \frac{(1+\epsilon){\tt g}}{4}} P_{2l}(x)
\,\, . ~~~~ &&  \label{facsols}
\end{eqnarray}
In the limit ${\tt g} = 1$ the $l=0$ term in the first series
diverges for $\epsilon = -1$. There are no divergences in the case
${\tt g}=3$, as we see from eq. (\ref{sols}). We postpone the
discussion of the limit ${\tt g}=1$ until the next section, and
for the remainder of this section take ${\tt g} \ne 1$.

Recalling the definition of the generating function for Legendre
polynomials, $(1 - 2x \tau + \tau^2)^{-1/2} = \sum_{l=0}^\infty
P_l(x) \tau^l$, and using
\begin{equation}
\lim_{\varepsilon\rightarrow 0} \int^\tau_\varepsilon d \vartheta
\vartheta^{l+\beta}= \frac{\tau^{l+\beta+1}}{l+\beta+1} -
\lim_{\varepsilon\rightarrow 0} \int^\varepsilon d \vartheta
\vartheta^{l+\beta} \label{regulator}
\end{equation}
to regulate divergences when $l + \beta \le -1$, we integrate the
generating function expansion over $\tau$. This yields the
identity
\begin{eqnarray}
&& \sum_{l=0}^\infty \frac{\tau^{2l+\beta+1}}{2l+\beta+1}
P_{2l}(x) =
\label{regdetails} \\
&&\frac12 \lim_{\varepsilon \rightarrow 0} \Bigl\{ 2
\sum_{l=0}^{l_{\max} \le -(1+\beta)/2} P_{2l}(x)
\int^{\varepsilon} d\vartheta \vartheta^{2l+\beta} + \nonumber \\
&& + \int^\tau_\varepsilon d\vartheta \vartheta^\beta \Bigl(
\frac{1}{\sqrt{1-2x\vartheta + \vartheta^2}} +
\frac{1}{\sqrt{1+2x\vartheta + \vartheta^2}} \Bigr) \Bigr\} \, .
\nonumber
\end{eqnarray}
Applying eq. (\ref{regdetails}) to eq. (\ref{facsols}), after
straightforward manipulations and performing explicitly the limit
$\varepsilon \rightarrow 0$ we find
\begin{eqnarray}
f(\Omega,z) = \frac{p}{2\pi M^2_4}
\frac{\tau^{\frac{1-3\epsilon}{2}}}{1 - \frac{1-3\epsilon}{4}{\tt
g}} -  \frac{p}{2(3 + \epsilon {\tt g}) \pi M^2_4} \times
~~~~~~~~ && \nonumber \\
\times \int^\tau_0 d\vartheta \Bigl(
\frac{1}{\sqrt{1-2x\vartheta+\vartheta^2}} +
\frac{1}{\sqrt{1+2x\vartheta+\vartheta^2}} -2 \Bigr)
~~~~~ && \label{intsols}\\
\times \Bigl( \frac{[3 - (1-\epsilon) {\tt g}]
\vartheta^{\frac{{\tt g}(1-\epsilon)-4}{2}}}{\tau^{\frac{({\tt
g}-3)(1-\epsilon)}{2}}} + \frac{[3 + (1+\epsilon){\tt
g}]\vartheta^{\frac{{\tt g}(1+\epsilon)+2}{2}}}{\tau^{\frac{({\tt
g}+3)(1+\epsilon)}{2}}} \Bigr) \, . \nonumber &&
\end{eqnarray}
When ${\tt g} \ne 1$ the integral (\ref{intsols}) is finite and
well-defined everywhere except at $x = \pm 1$, where it has the
usual short-distance singularities, because it is a Green's
function in the transverse spatial directions and $x = \pm 1$ is
where the sources are located. We remove the spurious poles in the
limit $\varepsilon \rightarrow 0$ in the integral representation
of (\ref{intsols}), arising from the implementation of the
regulator (\ref{regulator}), (\ref{regdetails}), by the
introduction of the factor $-2$ in the second line of the
integrand of (\ref{intsols}).

For completeness' sake, we note that the integral (\ref{intsols})
can be expressed in terms of the two-variable hypergeometric
functions $F_1(\alpha,\beta_1,\beta_2,\gamma;p,q)$
\cite{gradrhyz}, which are defined by the integral representation
\begin{eqnarray}
F_1(\alpha,\beta_1,\beta_2,\alpha+\mu;p,q) =
\frac{1}{B(\alpha,\mu)} \int_0^1 d\zeta \times ~~~~~~~~ && \nonumber \\
~~~~~~~ \times \zeta^{\alpha-1} (1-\zeta)^{\mu-1}
(1-p\zeta)^{-\beta_1} (1-q \zeta)^{-\beta_2} && \label{hyperg}
\end{eqnarray}
valid when $Re \, \alpha > 0$, $Re \, \mu > 0$. As an
illustration, we will only consider here the shock wave profile
$f$ on the brane, at $z=0$ (i.e. $\tau = 1$), on the
self-inflating branch $\epsilon = -1$. Similar derivation can also
be performed for the normal branch $\epsilon = 1$. We will split
up the integral in (\ref{intsols}) into a sum of two integrals,
with terms $\propto (1 \mp 2x \vartheta + \vartheta^2)^{-1/2} -
1$, and evaluate each integral separately by using the Euler
substitutions $1+ \sqrt{2(1\mp x)}\zeta = \sqrt{1\mp
2x\vartheta+\vartheta^2}+ \vartheta$. To see how the
hypergeometric functions in (\ref{hyperg}) emerge, consider e.g.
\begin{equation}
I = \int^1_0 d\vartheta
(\frac{1}{\sqrt{1-2x\vartheta+\vartheta^2}} -1 ) ( \frac{3 - 2
{\tt g}}{3-{\tt g}} \, \vartheta^{{\tt g}-2} + \frac{3}{3- {\tt
g}} \, \vartheta ) \, , \label{inthyps}
\end{equation}
and plug in $1+ \sqrt{2(1-x)} \zeta = \sqrt{1 -
2x\vartheta+\vartheta^2}+ \vartheta$. After straightforward
algebra, we can rewrite the integral (\ref{inthyps}), after
introducing $Q = -\sqrt{(1-x)/2} = -\sin (\theta/2)$, as
\begin{eqnarray}
I &=& \int^1_0 d\zeta \zeta \frac{x + Q \zeta}{(\zeta - Q)^2}
\Bigl\{ \frac{3 - 2 {\tt g}}{3- {\tt g}} \,
\Bigl(\frac{\zeta(1-Q\zeta)}{\zeta-Q}\Bigr)^{{\tt g}- 2}
\nonumber \\
&& ~~~~~~~~~~~~~~~~~~~~~~ + \frac{3}{3-{\tt g}} \,
\frac{\zeta(1-Q\zeta)}{\zeta-Q}  \Bigr\} \, . \label{intreps}
\end{eqnarray}
Breaking up the integrand into the powers of the kind that enter
in (\ref{hyperg}) we can rewrite (\ref{intreps}) as
\begin{eqnarray}
I &=& \frac{(3 - 2 {\tt g}) \Gamma({\tt g}-2)}{(3- {\tt
g})\Gamma({\tt g}-1)} \, \frac{x F_1({\tt g}-2,2-{\tt g}, {\tt g},
{\tt g}-1; Q, \frac{1}{Q})}{ (-Q)^{\tt g}} \nonumber
\\
&-& \frac{(3 - 2 {\tt g}) \Gamma({\tt g}-1)}{(3- {\tt
g})\Gamma({\tt g})} \, \frac{ F_1({\tt g}-1,2-{\tt g}, {\tt g},
{\tt g}; Q, \frac{1}{Q})}{(-Q)^{{\tt g}-1}} \nonumber
\\
&+& \frac{3}{(3- {\tt g})} \, \frac{x F_1(1,-1,3,2; Q,
\frac{1}{Q})}{ (-Q)^{3}} \nonumber \\
&-& \frac{3}{2(3- {\tt g})} \,  \frac{ F_1(2,-1,3,3; Q,
\frac{1}{Q})}{(-Q)^2}  \, . \label{intrephyps}
\end{eqnarray}
The remaining integrals in (\ref{intsols}) can be found from
(\ref{intrephyps}) by substituting $x \rightarrow -x$. Then, using
$P= -\sqrt{(1+x)/2} = -\cos (\theta/2)$, we finally obtain
\begin{eqnarray}
&& f(\Omega) = \frac{p}{2\pi M^2_4} \frac{1}{1-{\tt g}} -
\frac{p}{2\pi (3-{\tt g}) M^2_4} \Bigl\{ \nonumber
\\
&& \frac{(3 - 2 {\tt g}) \Gamma({\tt g}-2)}{\Gamma({\tt g}-1)} \,
\frac{x F_1({\tt g}-2,2-{\tt g}, {\tt g}, {\tt g}-1; Q,
\frac{1}{Q})}{ (-Q)^{\tt g}} \nonumber
\\
&-& \frac{(3 - 2 {\tt g}) \Gamma({\tt g}-1)}{\Gamma({\tt g})} \,
\frac{ F_1({\tt g}-1,2-{\tt g}, {\tt g}, {\tt g}; Q,
\frac{1}{Q})}{(-Q)^{{\tt g}-1}} \nonumber
\\
&+& 3 \Bigl(\frac{x F_1(1,-1,3,2; Q, \frac{1}{Q})}{ (-Q)^{3}} -
\frac{ F_1(2,-1,3,3; Q,
\frac{1}{Q})}{2 (-Q)^2} \Bigr) \nonumber \\
&-& \frac{(3 - 2 {\tt g}) \Gamma({\tt g}-2)}{\Gamma({\tt g}-1)} \,
\frac{x F_1({\tt g}-2,2-{\tt g}, {\tt g}, {\tt g}-1; P,
\frac{1}{P})}{ (-P)^{\tt g}} \nonumber
\\
&-& \frac{(3 - 2 {\tt g}) \Gamma({\tt g}-1)}{\Gamma({\tt g})} \,
\frac{ F_1({\tt g}-1,2-{\tt g}, {\tt g}, {\tt g}; P,
\frac{1}{P})}{(-P)^{{\tt g}-1}} \label{finsols}
\\
&-& 3 \Bigl(\frac{x F_1(1,-1,3,2; P, \frac{1}{P})}{ (-P)^{3}} +
\frac{ F_1(2,-1,3,3; P, \frac{1}{P})}{2 (-P)^2} \Bigr) \Bigr\} \,
. \nonumber
\end{eqnarray}
In the expressions for the hypergeometric functions one would
encounter an additional, spurious, logarithmic singularity. This
singularity is manifest in the integrals (\ref{inthyps}) or
(\ref{intreps}) in the lower limit of integration. However, after
Euler substitutions for integration variables, one has to
carefully relate the singularities in (\ref{inthyps}) and
(\ref{intreps}), recalling that these integrals are defined as
principal values of integration over $\vartheta$, in accordance
with the regularization prescription in (\ref{regulator}). After
this is done, these spurious logarithmic singularities cancel
precisely between the $Q$ and $P$ dependent terms in
(\ref{finsols}), leaving only the physical short distance
singularities carried by the null sources on the brane. We will
revisit this more closely in section V.

\section{Resonance On The Self-Inflating Branch}

As we have mentioned above, the solution (\ref{facsols}) for the
shock wave displays a spectacular behavior on the self-inflating
branch, $\epsilon = - 1$, as ${\tt g}\rightarrow 1$. In this limit
the tension vanishes, as can be seen from eq. (\ref{hubble}). The
$l=0$ term of the first sum has a pole there. No such poles arise
on the self-inflating branch, where the series (\ref{facsols}) is
finite. What then does this divergence mean? To gain some insight
we revisit the boundary value problem (\ref{bvprob}) for the $l=0$
mode but slightly relax the boundary conditions in the bulk, by
retaining the $l=0$ solution which is not square-integrable as
well. Writing this mode as
\begin{equation}
f_0 = A_0 e^{-2 H |z|} + B_0 e^{- H |z|} \, ,
\label{specasymptotic}
\end{equation}
and solving for $A_0$ and $B_0$ we find that when ${\tt g}=1$ the
mode $\propto A_0$ is an eigenmode of the homogeneous ($p=0$)
boundary value problem (\ref{bvprob}), completely cancelling on
the LHS of the last of eq. (\ref{bvprob}). This is the origin of
the divergence in (\ref{facsols}) for ${\tt g}=1$ and
mathematically is an example of {\it the Fredholm's alternative}:
either the inhomogeneous differential equation or the associated
homogeneous differential equation (i.e. with non-vanishing source
or with vanishing source, respectively) have a solution, but not
both at the same time. On the other hand, the mode $\propto B_0$
remains. Hence we could solve the last of (\ref{bvprob}) for it:
\begin{equation}
B_0 = \frac{p}{2\pi M^2_4} \, . \label{nonintegr}
\end{equation}
Because we have excluded the non-integrable bulk modes with $l>0$,
this mode is the fastest-growing one in the limit $z \rightarrow
\infty$ among all the allowed solutions of (\ref{bvprob}) for any
$l$. Thus far from the brane the resulting wave profile behaves as
\begin{equation}
f \rightarrow \frac{p}{\pi M_4^2} e^{- H |z|} \, .
\label{deepbulk}
\end{equation}
Hence as $z$ approaches the Cauchy horizon at $z \rightarrow
\infty$ (see Fig. \ref{fig:three}.) the asymptotic metric
(\ref{wave}) is
\begin{eqnarray}
ds_5^2 &\rightarrow& e^{-2 \epsilon H |z|} \Bigl\{ \frac{4
dudv}{(1+H^2 uv)^2}  + (\frac{1-H^2 uv}{1+H^2 uv})^2 \,
\frac{d\Omega_2}{H^2} \nonumber \\
&& ~~~~~~~~~~~~ + dz^2 \Bigr\} - \frac{4 p}{\pi M^2_4}
e^{-\epsilon H |z|} \delta(u) du^2 \, . \label{nonintwave}
\end{eqnarray}
Because the last term remains finite, the shock wave is not
localized to the brane, but extends deep into the bulk, as
anticipated. This solution carries a lot of energy and curvature
into the bulk, similarly to the gravitational pp-waves in the
Randall-Sundrum setup, studied in \cite{chagib}. Indeed, we can
rewrite the asymptotic metric (\ref{nonintwave}) expressed in
coordinates $u,v,z$ in terms of the original static patch
coordinates $r,t,w$ (see the text above eq. (\ref{nullback})), and
transform further to the spherical polar coordinates in the bulk
far from the brane, defined by
\begin{eqnarray}
w &=& \frac{1}{H} - \sqrt{\rho^2 - \tau^2} \, , \nonumber \\
r &=& \frac{1}{H} \frac{\rho}{\sqrt{\rho^2-\tau^2}} \sin \varphi \, , \nonumber \\
\sinh(Ht) &=& \frac{\tau}{\sqrt{\rho^2 \cos^2\varphi - \tau^2}} \,
. \label{flatcoords}
\end{eqnarray}
Then the asymptotic metric (\ref{nonintwave}) becomes
\begin{equation}
ds_5^2 \rightarrow -d\tau^2 + d\rho^2 + \rho^2 d\Omega_3 - \frac{8
p  \sqrt{\rho^2 - \tau^2} }{\pi M^2_4 H}\delta({\cal X}^2 - 1)
d{\cal X}^2 \, . ~~~ \label{nonintflat}
\end{equation}
where ${\cal X}= \frac{\rho \cos \varphi + \tau}{\sqrt{\rho^2 -
\tau^2}+\rho \sin \varphi}$, and $d\Omega_3 = d\varphi^2 +
\sin^2\varphi d\Omega_2$ is the line element on a unit 3-sphere.
Since this asymptotic metric is invariant under $S^2$ rotations,
the solution looks like an imploding shell of energy in Minkowski
space. For a hint of how such a shell is generated, move away from
${\tt g=1}$ and reconsider the boundary value problem
(\ref{bvprob}) while still keeping the non-integrable bulk wave
function in the $l=0$ mode, as in eq. (\ref{specasymptotic}). Then
(\ref{bvprob}) relates $A_0$ and $B_0$ according to
\begin{equation}
B_0 = \frac{1}{2-{\tt g}} \Bigl(\frac{p}{2\pi M^2_4} - 2 (1-{\tt
g}) A_0 \Bigr) \, . \label{relcoeff}
\end{equation}
Thus we see that when ${\tt g} \ne 1$, the localized mode $\propto
A_0$ can absorb fully the effect of the source $\propto p$. This
is energetically favorable, because the non-integrable mode
$\propto B_0$ decouples because it has vanishing overlap with the
brane where the source is located. Hence we can set $B_0 = 0$ and
forget about it. However in the limit ${\tt g} = 1$ the
modifications of gravity occur at the scale equal to the
cosmological horizon on the brane. Brane and bulk begin to
``resonate": in a time-dependent problem, as ${\tt g}\rightarrow
1$ slowly, the localized $l=0$ mode would grow ever larger,
spreading further out into the bulk. It would develop a larger
overlap with the bulk graviton, so that it can produce delocalized
gravitons in order to shed the energy which drives its unbounded
growth. In this way as ${\tt g} \rightarrow 1$ energy could escape
into the bulk. This indicates an onset of a dramatic new
instability, which warrants further investigation.


\section{Shock Waves at Short Distances}

In this section we consider $f$ on the brane, when $z=0$ (i.e.
$\tau = 1$), at transverse distances ${\cal R}$ well inside the
cosmological horizon, ${\cal R } \ll H^{-1}$. Note that we can
rewrite eq. (\ref{hubble}) as $(1 + \epsilon {\tt g}) H^2 =
\lambda/(3M^2_4)$. This shows that on the normal branch ${\tt g}$
can vary between zero, in which case we recover the 4D limit, and
infinity, when we recover the 5D limit, with the ``resonance"
discussed in detail in the previous section at ${\tt g}=1$. On the
other hand, on the self-inflating branch, ${\tt g} \le 1$ or $r_c
\ge 1/H$, as long as $\lambda \ge 0$. In the limit ${\tt
g}\rightarrow 1$ on the self-inflating brane, as mentioned before,
the tension vanishes.

Let us start with ${\tt g} = 0$. In this case (\ref{intsols}) is
identical to the 4D case on both branches. Indeed, it can be
expressed as $f = \frac{p}{2\pi M^2_4}( 1 - I_- - I_+)$ where
$I_\mp$ are the integrals
\begin{equation}
I_\mp  = \int^1_0 d\vartheta (\frac{1}{\vartheta^2}+\vartheta)
\Bigl(\frac{1}{\sqrt{1\mp 2x \vartheta + \vartheta^2}} - 1 \Bigr)
\, , \label{integs}
\end{equation}
which obey $I_+(x) = I_-(-x)$. Hence as in section III it is
enough to evaluate one, say $I_-$. Using the Euler substitution
$\xi = \sqrt{1 - 2x\vartheta+\vartheta^2}+ \vartheta$ we can
rewrite the integral $I_-$ as
\begin{eqnarray}
I_- &=& 4x \int_{1+\varepsilon_-}^{1+\sqrt{2(1-x)}}
\frac{d\xi}{(\xi+1)^2
(\xi-1)} \nonumber \\
&-&  2  \int_{1}^{1+\sqrt{2(1-x)}}
\frac{d\xi}{(\xi+1)^2}
\nonumber \\
&-&  \frac14 \int_1^{1+\sqrt{2(1-x)}} d \xi \frac{(z-1)^2
(z+1)}{(\xi-x)^2} \nonumber \\
&+& \frac{1+x}4 \int_1^{1+\sqrt{2(1-x)}} d \xi \frac{(z-1)^2
(z+1)}{(\xi-x)^3} \, , \label{splitints}
\end{eqnarray}
where in the first line we have replaced $1 \rightarrow
1+\varepsilon_-$ in the lower limit of integration in order to
regulate a spurious logarithmic singularity, which we have
discussed in section III. A direct evaluation of the integrals
yields
\begin{equation}
I_- = x \ln \frac2{\varepsilon_-} - x - \frac12 \, ,
\label{intminus}
\end{equation}
and thus, using $I_+(x) = I_-(-x)$, and noting that the regulator
is now $\varepsilon_+$, also
\begin{equation}
I_+ = - x \ln \frac2{\varepsilon_+} + x - \frac12 \, ,
\label{intplus}
\end{equation}
such that the final result for $I = I_++I_-$ is
\begin{equation}
I = x \ln\frac{\varepsilon_+}{\varepsilon_-} - 1 \, .
\label{intsum}
\end{equation}
To determine the ratio of the regulators
$\frac{\varepsilon_+}{\varepsilon_-}$, recall that they are the
image of the regulator $\varepsilon$ for the variable $\vartheta$
under the Euler substitution for the integration variables in
$I_\pm$. That yields $\varepsilon_\mp = (1 \mp x) \varepsilon
\label{regrels}$, so that
\begin{equation}
\frac{\varepsilon_+}{\varepsilon_-} = \frac{1+x}{1-x} \, .
\label{regratio}
\end{equation}
Therefore the singularities indeed cancel out, and
\begin{equation}
I = x \ln[\frac{1+x}{1-x}] - 1 \, . \label{intsumfin}
\end{equation}
Substituting this in the equation for the shock wave profile
finally yields
\begin{equation}
f_{4D}(\Omega) = \frac{p}{2 \pi M^2_4} \Bigl( 2 - x \,
\ln[\frac{1+x}{1-x}] \Bigr) \, . \label{4dsols}
\end{equation}
The metric transverse to the null particle on the brane is
$ds_2^2|_{z=u=0} = d\Omega_2/H^2$. Therefore the proper ``radial"
transverse distance for small polar angles $\theta$ is ${\cal R}
\simeq \theta/H$, so that $x = 1- H^2 {\cal R}^2/2$. Thus the
shock wave profile (\ref{4dsols}) reduces precisely to the flat 4D
solution \cite{aichsexl,thooft}: up to ${\cal O}({\cal
R}^2/H^{-2})$ corrections, using the sign conventions of
\cite{hota,kostas}, we find
\begin{equation}
f_{4D}(\Omega) = \frac{p}{\pi M^2_4} + \frac{p}{\pi M^2_4}
\ln(\frac{\cal R}{2H^{-1}}) \, . \label{4dsolflat}
\end{equation}
As long as it is finite, the numerical value of the constant term
in this equation is physically unimportant since a finite constant
term in the shock wave profile in a flat 4D spacetime can always
be changed by a simple 4D diffeomorphism \cite{thooft}. We cannot
write down the integrals which incorporate the ${\cal O}({\tt g})$
corrections in closed form because they contain terms like $\int
d\zeta \ln(1+b\zeta)/\zeta$ \cite{gradrhyz}. Nevertheless, it is
easy to extract their short-distance effects: at short distances
the singular ${\cal O}({\tt g})$ powers in the integrand precisely
cancel out. In the remainder, since the leading order singularity
in (\ref{intsols}) is logarithmic around ${\tt g}=0$, the
additional logarithms soften the corrections further, and render
them finite as ${\cal R} \rightarrow 0$. Hence at short distances
the solution looks exactly the same as in 4D, and the corrections
are negligible when ${\cal R} < H^{-1}/{\tt g}$.

In fact this behavior persists for any finite value of ${\tt g}$.
We can see this from identifying the leading contribution in
(\ref{intsols}) at short distances. Rewriting the integrals
(\ref{intsols}) by using the same substitutions $\xi = \sqrt{1\mp
2x\vartheta+\vartheta^2}+ \vartheta$ and looking at the limit $1-x
\ll 1$, we identify the leading singularity. After cancelling the
spurious terms as above, for all values of ${\tt g} \ne 1$ and
$\epsilon$ the leading singularity at short distances comes from
the term
\begin{equation}
- \frac{p}{\pi M^2_4} \int_1^{1+\sqrt{2(1-x)}} \frac{d\xi}{\xi-x}
= \frac{p}{\pi M^2_4} \ln\sqrt{\frac{1-x}{2}} + {\rm finite} \, .
\label{sings}
\end{equation}
Substituting $x = 1- H^2 {\cal R}^2/2$, we recover the leading
logarithm in (\ref{4dsolflat}). The corrections to this result at
larger distances in general differ from the corrections which
arise when one expands the 4D de Sitter solution (\ref{4dsols})
around the short distance limit describing shock waves in
asymptotically flat space (\ref{4dsolflat}), where one finds
corrections to come as even powers of $H{\cal R}$. In DGP,
however, the corrections in (\ref{logs}) will arise as odd powers
of $H{\cal R}$ as well, starting with the linear term. This
signals the presence of the fifth dimension concealed by the
``gravity filter". Therefore in general we will have
\begin{eqnarray}
f(\Omega) &=& \frac{p}{\pi M^2_4} \Bigl((1 + a_1 H^2 {\cal R}^2 +
...)
 \ln(\frac{\cal R}{2H^{-1}}) \nonumber \\
&&+ {\rm const} + b_1 H {\cal R} + b_2 H^2 {\cal R}^2 + ... \Bigr)
\, ,
 \label{logs}
\end{eqnarray}
where the coefficients $a_k, b_k$ are numbers that can be computed
explicitly for given values of ${\tt g}$ and $\epsilon$, and we
have seen that e.g. $b_1 \sim {\tt g}$.

To see how 5D gravity reemerges, we can take the limit $g
\rightarrow \infty$ on the normal branch $\epsilon = 1$, keeping
$M_5$ fixed. This removes the brane-localized terms $\propto M_4$
from the action and neutralizes the ``gravity filter". In this
case the background solution reduces to the inflating brane in 5D
Minkowski bulk \cite{kalinde}. From eq. (\ref{facsols}) we see
that the shock wave profile on the brane at $z=0$ in the limit $g
\rightarrow \infty$ reduces to
\begin{equation}
f_{5D}(\Omega) = - \frac{p H}{ \pi M^3_5 } \sum_{l=0}^\infty
P_{2l}(x) - \frac{3 p H }{2 \pi M^3_5} \sum_{l=0}^\infty
\frac{P_{2l}(x)}{2l-1}  \, , \label{5dsols}
\end{equation}
where we have used $M^2_4 {\tt g} = 2M^3_5/H$. We can evaluate the
first sum using the identity
\begin{equation}
\sum_{l=0}^\infty P_{2l}(x) = \frac{1}{\sqrt{8(1-x)}} +
\frac{1}{\sqrt{8(1+x)}} \, , \label{identity}
\end{equation}
and reduce the second sum to the integral representation
\begin{eqnarray} \sum_{l=0}^\infty \frac{P_{2l}(x)}{2l-1} &=& \frac12 \int^1_0
\frac{d\vartheta}{\vartheta^2}
\Bigl(\frac{\vartheta}{\sqrt{1-2x\vartheta + \vartheta^2}}
\nonumber \\
&& ~~~ + \frac{\vartheta}{\sqrt{1+2x\vartheta +\vartheta^2}} - 2
\Bigr) - 1 \, . \label{inteqssec} \end{eqnarray}
Evaluating the integrals in much the same way as above, and
substituting back in (\ref{5dsols}) we find
\begin{eqnarray}
&&f_{5D}(\Omega) = - \frac{p H}{2\sqrt{2} \pi
M^3_5}\Bigl(\frac1{\sqrt{1-x}} + \frac{1}{\sqrt{1+x}} \Bigr)
\nonumber \\
&& ~~~~~~~ + \frac{3pH}{4\pi M^3_5} \, x \,
\ln\Bigl[\frac{\sqrt{\frac{1-x}2} (1+\sqrt{\frac{1-x}2})
}{\sqrt{\frac{1+x}2}(1+\sqrt{\frac{1+x}2})}\Bigr] \nonumber \\
&& ~~~~~~~  + \frac{3pH}{2\pi M^3_5} \Bigl(\sqrt{\frac{1-x}2} +
\sqrt{\frac{1+x}2} \Bigr) \, . \label{5dresults}
\end{eqnarray}
At short transverse distances $x = 1- H^2 {\cal R}^2/2$\, ignoring
a constant and higher powers of ${\cal R}$, we find the leading
order behavior of $f_{5D}$ to be
\begin{equation}
f_{5D}(\Omega) = - \frac{p}{2 \pi M^3_5 {\cal R}} + \frac{3p
H}{4\pi M^3_5} \ln(\frac{\cal R}{2H^{-1}})  \, . \label{5dlims}
\end{equation}
The first term is precisely the $5D$ shock wave solution
considered in \cite{gabriele,devega}. The second term arises from
the residual 4D graviton zero mode which persists on the normal
branch in the limit ${\tt g} \rightarrow \infty$ because the bulk
volume remains finite.
Its perturbative coupling is
set by the effective 4D Planck scale $M_{4 \, eff}^2$, which can
be computed from the normalization of the zero mode graviton wave
function in the bulk \cite{conformal,rs2} as
\begin{equation}
M_{4 \, eff}^2 = 2 M^3_5 \int^\infty_0 dz \, \Omega^{3/2} =
\frac{4M^3_5}{3H} \, . \label{4dplancks}
\end{equation}
Substituting this in the formula for the 4D shock wave profile at
short distances (\ref{4dsolflat}) exactly reproduces the second
term in (\ref{5dlims}). Notice the relative sign difference
between the two terms on the RHS of (\ref{5dlims}). This is
necessary in order for both to yield an {\it attractive} force $
\propto - \vec \nabla f$ on a test particle in the shock wave
background. Therefore the result (\ref{5dlims}) precisely meets
the expectations for the shock wave profile arising in a theory
with both a 5D and 4D graviton modes in flat bulk. Because the two
gravitons are equally coupled, the theory looks effectively
five-dimensional at all sub-horizon distances, where the inverse
power of ${\cal R}$ in (\ref{5dlims}) wins over the logarithm.
Hence indeed in this limit the effects of the 4D graviton are
completely negligible inside the cosmological horizon, and the
``gravity filter" has been turned off.

This analysis clearly demonstrates how the ``gravity filter" of
\cite{dgp} works even beyond perturbation theory. It is encoded in
the behavior of the coefficients in (\ref{facsols}), which
decrease with the index $l$ of the Legendre polynomials. Because
the graviton momentum on a transverse 2-sphere along de Sitter
brane is proportional to the index of the polynomial, $q \sim H
l$, the modes with $q > {\tt g} H \simeq 1/r_c$ have amplitudes
suppressed by $q$. Hence the short-distance divergences of the
series can be no worse than a logarithm for any finite value of
${\tt g}$. Indeed, at short distances ${\cal R} \le r_c \simeq
H^{-1}/ {\tt g}$, our shock waves are exactly the same as in 4D
GR. For very low momenta the amplitudes are not suppressed, and so
the shock wave profile changes towards 5D behavior at large
distances ${\cal R} \ge H^{-1}/ {\tt g}$. The limit ${\tt g}
\rightarrow \infty$ on the normal branch illustrates this
explicitly: when $r_c \rightarrow 0$ all modes with finite momenta
remain unsuppressed and we recover 5D gravity.

\section{Planckian Scattering}

In this section we explore the scattering of two relativistic
particles on the brane, $m_k/p_k \ll 1$, whose energy-momenta
$p_1, p_2$ reach beyond the Planck scale. In this regime, gravity
will generically dominate over other interactions. Indeed in 4D
the effective coupling of gravitons to the packets of
stress-energy is $\alpha \simeq G_N s$, where $s = (p_1-p_2)^2 =
4E^2$ is the Mandelstam $s$-parameter and $E$ the center-of-mass
(COM) energy \cite{thoofteikon,muso,acive}. It has been shown that
even if it exceeds the Planck scale, the scattering cross section
can still be reliably calculated using the methods of quantum
field theory and GR \cite{thoofteikon,muso,acive,verlinde,more}.
Classical gravitational scattering has been studied before by
D'Eath and Payne \cite{deathpayne}, and the eikonal approximation
calculations of \cite{thoofteikon,muso,acive} are consistent with
these results when the impact parameter $b$ is larger than the
center-of-mass Schwarzschild radius $\sim G_N \sqrt{s}$. The
problem of Planckian scattering and black hole formation has been
revived recently in the context of braneworld constructions, since
if the fundamental scale of gravity is really low, it may be
possible to form black holes in high energy processes in colliders
\cite{bafi,dila,gito,giuraw,aref,roberto,eagi}. Thus one might
wonder if such processes would be different in interesting ways in
the framework of brane-induced gravity.

However, in light of our findings above, it is clear that in the
case of DGP with finite ${\tt g}$, when the impact parameter $b$
is smaller than the scale of modification of gravity, $b <
H^{-1}/{\tt g}$, Planckian gravitational scattering will be very
well approximated by the conventional 4D GR description. The
differences are suppressed by the powers of $m/p$ and $H{\cal R}$,
which are both very small for highly energetic particles at short
distances.

Let us now explicitly verify this. In the limit when the impact
parameter is smaller than the DGP scale of modification of
gravity, $b < H^{-1}/{\tt g}$, the scattering is mediated
predominantly by the 5D modes with transverse momenta $q > 1/b >
{\tt g} H$. These individual 5D modes couple to a brane-localized
source with the effective coupling
\begin{equation}
G_{N \, eff}(q) \sim \frac{1}{M^2_4}\frac{H}{q} \, ,
\label{effcoupl} \end{equation}
as can be seen directly from
(\ref{facsols}), and is consistent with \cite{dgp}. The latter
term is the ``gravity filter", which cuts off the influence of the
modes with high transverse momentum. This is in contrast to what
happens in the Randall-Sundrum models \cite{rs2}, where the modes
carrying greater momentum compete more efficiently with the
zero-mode graviton \cite{roberto}. The couplings can be rewritten
as
\begin{equation}
G_{N\, eff}(q) \sim \frac{1}{2M^3_5} \, \frac{1}{{\tt g} r_c} \,
\frac1{r_c q} \, , \label{5dcoupls} \end{equation}
where the first factor is the usual $5D$ bulk coupling, with the
factor of $2$ accounting for the orbifolding, the second factor is
the ``volume dilution" $\sim 1/({\tt g} r_c) = H$, and the last
factor is the ``filter". Hence in order for a mode mediating a
process on the brane with a transverse momentum $q$ to be strongly
coupled, the process should unravel at COM energies $E$ where
$G_{N \, eff}(q) E^2 \ge 1$, or therefore when $E  \ge
\sqrt{\frac{q}{H}} M_4$. This should be compared with the
requirements for the processes in the bulk, where the filter does
not operate, because they are mediated by delocalized modes, which
become strongly coupled at much lower scales $E \ge M_5$.

Therefore we see that for the brane processes at energies $
M_4/\sqrt{{bH}} \ge E \ge M_4$, all the individual modes with
momenta $q >1/b$ are in fact weakly coupled, with $G_{N\,
eff}(q)<1$. However, the {\it resonance} mimicking the 4D graviton
is {\it strongly} coupled. Indeed, this resonance couples to the
brane-localized sources with the effective strength
\begin{equation}
G_N \sim \sum_{q \ge 1/b} G_{N\, eff}(q) \sim
\frac{1}{M^2_4}\int^\Lambda_{1/b} \frac{dq}{q} = \frac{1}{M^2_4}
\ln(\Lambda b)\, , \label{resoncoupl}
\end{equation}
where the UV cutoff $\Lambda$ is the highest transverse momentum
for which the particle scattering is meaningful, which we roughly
estimate as $\Lambda \sim M_4$. Thus $G_N \sim \frac{1}{M^2_4}$ up
to logarithmic corrections, and so at COM energies $E> M_4$ indeed
$G_N E^2 > 1$, implying that the resonance-mediated scattering is
in the strong coupling regime, where it should dominate over other
processes much like the conventional 4D gravity. This description
should be reasonable as long as the impact parameter $b$ is
greater than the Schwarzschild radius $\sim G_{N} E \simeq G_N
\sqrt{s}$ \cite{schwarz}, where black holes are expected to form
\cite{thoofteikon,acive,deathpayne,eagi}, and where one would need
to go beyond the eikonal approximation. It is interesting however
that in the regime of COM energies $M_4 \le E \le M_4/\sqrt{bH}$,
and impact parameters $b \ge G_N E$, which yield the COM energy
interval $M_4 \le E \le M_4 (M_4/H)^{1/3}$, the strongly coupled
4D graviton resonance is resolved into the individual weakly
coupled bulk modes.

In this limit, the scattering amplitude can be extracted directly
from the shock wave solutions \cite{thoofteikon}. For a shock wave
profile $f(\Omega)$ on the brane at $z=0$, the formula relating
the scattering amplitude in the eikonal approximation ${\cal
A}(s,b)$, expressed via the eikonal ${\delta}_E(s,b) = \ln\Bigl(
\frac{{\cal A}(s,b)}{2i} \Bigr)$, and the shock wave metric
(\ref{wave}) is \cite{thoofteikon,roberto}
\begin{equation}
g_{uu} = \frac{p}{\pi} \, \delta(u) \, \frac{{\delta}_E(s,b)}{s}
\, . \label{eikonalform}
\end{equation}
Solving for ${\delta}_E(s,b)$ and substituting in the shock wave
profile formula from (\ref{wave}) and (\ref{logs}), we obtain,
taking into account our normalizations,
\begin{eqnarray}
{\delta}_E(s,b) &=& - \frac{s}{ M_4^2}
\Bigl((1 + a_1 H^2 { b}^2 + ...) \ln(\frac{ b}{2H^{-1}}) \nonumber \\
&&~~~~ + {\rm const} + b_1 H { b} + b_2 H^2 { b}^2 + ...\Bigr) \,
, \label{eikonform}
\end{eqnarray}
where $a_k, b_k$ are the same numerical coefficients as in eq.
(\ref{logs}) (and $b_1 \sim {\tt g}$). Thus clearly as long as $b
\ll r_c = H^{-1}/{\tt g}$, the deviations away from the expression
in 4D GR are completely negligible. Hence as we have claimed
above, the scattering will be very well approximated by the
analysis in the conventional 4D theory
\cite{thoofteikon,muso,acive}. As a consequence one expects also
that the processes of black hole formation in high energy
collisions would behave as in 4D GR \cite{bafi,eagi,deathpayne},
controlled by the low energy value of the Planck scale, $M_4 \sim
10^{19}$ GeV, which is much too high to see such processes in
forthcoming experiments at the LHC. This provides further
illustration of the efficiency of the ``gravity filter"
\cite{dgp}.

\section{Conclusions}

In this paper we have given a detailed derivation and analysis of
the exact gravitational shock wave metric in DGP models
\cite{dgp}, first presented in our earlier paper \cite{kallet}.
The shock wave solutions provide an explicit and concrete example
of the ``gravity filter" mechanism of \cite{dgp} beyond
perturbation theory. Even on these exact solutions, the effects of
the whole continuum of bulk graviton modes is dominated by a long
range resonance, which mimics the standard 4D graviton out to
distances $r_c = H^{-1}/{\tt g}$. This is because the bulk modes
which can transfer large transverse momentum on the brane $q> {\tt
g} H \simeq 1/r_c$ have amplitudes suppressed by $q$. Hence at
short distances ${\cal R} \le r_c$ their influence is cut off, and
the rate of divergence of the shock wave metrics can be no worse
than a logarithm for any finite value of ${\tt g}$, so that to the
leading order the gravitational shock waves are exactly the same
as in 4D GR. For very low momenta the amplitudes are not
suppressed, leading to IR corrections which change the shock wave
towards the 5D form at large distances ${\cal R} \ge H^{-1}/ {\tt
g}$. We have also demonstrated that the leading-order Planckian
scattering behaves in exactly the same way as in 4D GR. The
differences come with the powers of $m/p$ and $H{\cal R}$, and are
very small for highly energetic particles at short distances.
Further, we have noted a spectacular new channel for energy loss
into the bulk on the self-inflating branch, which arises when
gravity is modified at exactly the brane de Sitter radius.

We have argued earlier \cite{kallet} that our exact shock waves
may also provide for a new arena to explore the scalar graviton
sector in DGP. In the relativistic limit pursued here and in
\cite{kallet} the scalar graviton has decoupled, in spite of the
strong coupling issues that were encountered in na\"ive
perturbation theory \cite{strongcouplings,lpr,giga,nira}. However
one can imagine taking a very fast, but not ultra-relativistic
observer who explores the field of a non-vanishing source mass. In
this limit, one may treat the rest mass of the source as a
perturbation of the shock wave geometry, and study how it sources
the scalar graviton, and what happens to the graviton perturbation
theory around the shock wave. This could reveal new aspects of
modified gravitational physics in the IR, and help illuminate the
status of effective 4D theory in DGP.

\vskip.3cm

\smallskip

{\bf \noindent Acknowledgements}

\smallskip

\vskip.3cm

We thank S. Dimopoulos, G. Dvali, R. Emparan, G. Gabadadze, M.
Luty, K. Sfetsos and L. Sorbo for useful discussions, and the
Aspen Center for Physics for hospitality during the initiation of
this work. NK was supported in part by the DOE Grant
DE-FG03-91ER40674, in part by the NSF Grant PHY-0332258 and in
part by a Research Innovation Award from the Research Corporation.

\end{document}